\documentstyle[preprint,aps]{revtex}
\begin{document}
\draft
\preprint{} 
\title{ 
Farey Tree and the Frenkel-Kontorova Model}
\author{Hsien-chung Kao, Shih-Chang Lee, and Wen-Jer Tzeng}
\address{Institute of Physics, Academia Sinica, Taipei, Taiwan 11529, R.O.C.}
\maketitle
\begin{abstract}
We solved the Frenkel-Kontorova model with the potential $V(u)= -\frac{1}{2} 
|\lambda|(u-{\rm Int}[u]-\frac{1}{2})^2$ exactly.  For given $|\lambda|$,
there exists a positive integer $q_c$ such that for almost all values of the
tensile force $\sigma$, the winding number $\omega$ of the ground state 
configuration is a rational number in the $q_c$-th level Farey tree.  For fixed
$\omega=p/q$, there is a critical $\lambda_c$ when a first order phase 
transition occurs.  This phase transition can be understood as the 
dissociation of a large molecule into two smaller ones in a manner dictated 
by the Farey tree.  A kind of ``commensurate-incommensurate'' transition 
occurs at critical values of $\sigma$ when two sizes of molecules co-exist.
``Soliton'' in the usual sense does not exist but induces a transformation
of one size of molecules into the other.
\end{abstract}
\pacs{PACS number(s):03.20.+i, 05.45.+b, 64.60.-i}

%
\section{Introduction}  
Frenkel-Kontorova (FK) model describe a system of particles moving in an 
infinite sequence of potential wells. In the limit of shallow wells, the 
particles are kept at an equilibrium distance by a tensile force $\sigma$. 
Such models are also widely studied  in the context of circle maps when 
certain periodicity exists in the locations of the potential 
wells\cite{mm87}.  A stationary configuration in the FK model corresponds
to an orbit in the circle map.  In the study of FK models, however, one is 
particularly concerned with the minimum energy configurations.  The ground
state is a ``recurrent'' minimum energy configuration\cite{aub80i,aub83}
that is characterized by a winding number $\omega$, the inverse of which,
$1/\omega$, gives the average number of particles per well.  Hence a ground
state with rational $\omega=p/q$\cite{irreducible} consists of molecules that
are composed of $q$ particles and $p$ wells.

It is well-known that two types of phase transitions occur quite generally
in FK models\cite{gri90}.  The first type is the commensurate-incommensurate
transition which occurs at critical values of $\sigma$ when the enthalpy for
the creation of solitons or anti-solitons vanishes\cite{gri90}.  The second
type occurs at critical values of $\lambda$ characterizing the strength of 
the potential wells and corresponds, in the language of circle map, to the 
breaking of an invariant circle into a cantorus.  In the ground state with 
an irrational winding number $\omega$, the allowed positions of the particles
in the potential wells change from being the entire period in the 
``unpinned'' phase to being a cantor set in the ``pinned'' phase.  
Correspondingly, the hull function $f_{\omega}(x)$, from which one obtains
the position $u_n$ of the $n$-th particle by $u_n= f_{\omega}(n\omega)$, 
turns from a smooth function into a 
no-where-differentiable function\cite{aub80i,aub83,aub78,aub80,aub91}. 

These two types of phase transitions, however, are not the only phase 
transitions that could occur in a FK model.  In particular, when the 
potential possess an ``internal structure'', e.g., having multiple local
minima in a period, one will encounter first order phase transitions
at critical values of the height of a local minimum\cite{gsu90,klt96}. 
At the critical point, the filling fraction of the corresponding well
(the local minimum) changes from a value sub-commensurate to the winding
number $\omega$ to a value non-sub-commensurate to $\omega$\cite{klt96}.  As
in the case of commensurate-incommensurate transition, this new type of first
order transition is nucleated by solitons\cite{gsu90}.

In this letter, we would like to present yet a new type of first order
phase transition by solving exactly an FK model with a potential which is
concave almost everywhere.  The phase transition is not nucleated by 
solitons.  Rather, it can be understood as the breaking up of a large 
molecule into two smaller ones, in a manner dictated by the Farey tree.

The stationary configuration of an FK model is determined by the equation
\begin{equation}
u_{n+1}-2u_n+u_{n-1}=\lambda V'(u_n),\label{eq:forcebal}
\end{equation} 
where $u_n$ denotes the coordinate of the $n$-th particle.  The model we
consider has the potential 
\begin{equation}
V(u)=\frac{1}{2}\lambda\left(u-{\rm Int}[u]-\frac{1}{2}\right)^2,~~~~~
-4<\lambda<0, \label{eq:v(u)}
\end{equation} 
where ${\rm Int}[u]$ equals the largest integer not larger than $u$.  The 
case of $\lambda > 0$ had been thoroughly investigated before by several
authors\cite{klt95}.

The recurrent stationary configurations are characterized by a winding 
number $\omega$ and can be specified by a hull function $f_{\omega}(x)$
so that 
\begin{equation}
u_n=f_{\omega}(n\omega+\alpha),\label{eq:hull}   
\end{equation}
where $\alpha$ is a constant phase.  The hull function satisfies 
$f_{\omega}(x+1)=f_{\omega}(x)+1$.

For $\omega=p/q$, the hull function is a step function with $q$ steps in a 
period.  Depending on whether $f_{\omega}(x)$ is continuous at integer values
of $x$, there are two solutions:
\begin{equation}
f_{\omega}(x)=\sum_{n=0}^{q-1}\nu_n(q)\left({\rm Int}[x+n\omega ]
-\overline{\rm Int}[n\omega]\right), \label{eq:f(x)}
\end{equation}
and
\begin{equation}
\bar{f}_{\omega}(x)=[f_{\omega}(x+\frac{1}{2q})+f_{\omega}(x-
\frac{1}{2q})]/2, \label{eq:bf(x)},
\end{equation}
where $\nu_n(q)=\tan\frac{\chi}{2}\csc\frac{q\chi}{2}\cos(n-
\frac{q}{2})\chi$, $\chi=\arccos(1-|\lambda|/2)$ and $\overline{\rm Int}_[x] 
\equiv ({\rm Int}[x^+]+{\rm Int}[x^-])/2$.  The configuration described by 
$\bar{f}_{\omega}(x)$ has one particle sitting at the potential bottom in 
a given period.  $\bar{f}_{\omega}(x)$ can also be expressed as
\begin{equation}
\bar{f}_{\omega}(x)=\sum_{n=0}^{q-1}\bar{\nu}_n(q)\left({\rm Int}[x+n\omega 
+\frac{1}{2q}]-{\rm Int}[n\omega+\frac{1}{2q}]\right), \label{eq:bf(x)1}
\end{equation}
where  $\bar{\nu}_n=(\nu_n+\nu_{n'})/2$ with $n'\equiv n+q_1\pmod{q},~~
0\leq n'\leq q-1$, and $q_1$ is determined by the Farey tree in the 
following way.  Given the irreducible fraction $\omega=p/q$, $\omega_1=p_1/
q_1$ and $\omega_2=p_2/q_2$ is the unique irreducible pair such 
that\cite{farey}
\begin{equation}
p=p_1+p_2,~~q=q_1+q_2,~~\omega_1<\omega<\omega_2
\end{equation}
and 
\begin{equation} 
pq_1=p_1q+1,~~p_2q=pq_2+1,~~p_2q_1=p_1q_2+1. \label{eq:farq}
\end{equation}

We shall call the ``$q$-th level Farey tree'' to be the Farey tree where 
irreducible fractions with denominator larger than $q$ are truncated.  The
average energy per particle corresponding to the stationary configurations
described by $f_{\omega}(x)$ and $\bar{f}_{\omega}(x)$ are respectively
\begin{equation}
\Psi(\omega)=\frac{\omega^2}{2}-\frac{|\lambda|}{4}\sum_{n=0}^{q-1}\nu_n(q)
\left\{\frac{1}{4}-(n\omega-\frac{1}{2}-{\rm Int}[n\omega])^2\right\}
\label{eq:eom}
\end{equation}
and 
\begin{equation}
\bar{\Psi}(\omega)=\Psi(\omega)-\frac{|\lambda|}{8q}\nu_0(q).\label{eq:ediff}
\end{equation}
One notices that at $\chi=\chi_q\equiv \pi/q$, $\nu_0(q)$ vanishes and the
two configurations become degenerate.  Eq.~(\ref{eq:ediff}) also indicates
that for $0<\chi<\chi_q$, the ground state configuration with winding number
$\omega=p/q$ is given by $\bar{f}_{\omega}(x)$.

It is instructive to learn how the configuration described by $f_{\omega}(x)$
evolves as $\chi$ increases from zero.  Let $u_n=f_{\omega}(n\omega-1/2q)$
for $0\leq n\leq q-1$.  As $\chi$ increases from zero to $\chi_q$, the
$\nu_n$'s change from $\nu_0= \nu_1 = \cdots =\nu_{q-1} =1/q$ to a first zero
value occurring at $\nu_0$.  When this happens, the two particles 
$u_0$ and $u_{q_1}$ touch the bottom of the potential wells.  As $\chi$  
increases further, these two particles are pinned at the bottom and we can no
longer use $f_{\omega}(x)$, which ceases to be an increasing function, to 
describe this configuration.  Instead, the $q_1$ particles from $u_0$ to 
$u_{q_1-1}$ are described by an ``$\bar{f}_{\omega_1}$ section'', i.e., a 
consecutive $q_1$ particles chain with the first particle sitting at the 
potential minimum in the $\bar{f}_{\omega_1}$ configuration, while the $q_2$
particles from  $u_{q_1}$ to $u_{q_-1}$ are described by an 
$\bar{f}_{\omega_2}$ section.  For a general $\omega=p/q$, if we regard an 
$\bar{f}_{\omega}$ section as a ``molecule'' of size $(q,p)$, i.e., composed
of $q$ particles and $p$ wells, then $\chi_q$ can be seen to be a critical
point of a first order transition when a molecule of size $(q,p)$ is just 
about to break up into two molecules of sizes $(q_1,p_1)$ and $(q_2,p_2)$
respectively.  There three type of molecules, whose corresponding $\omega_1$,
$\omega$, and $\omega_2$ are related as consecutive Farey fractions, co-exist
at the critical point.  

For $\chi > \chi_q$, $\bar{f}_{\omega}(x)$ ceases to describe the ground 
state configuration with winding number $\omega$.  One can show from 
Eqs.~(\ref{eq:eom}) and (\ref{eq:ediff}) that
\begin{equation}
q_1\bar{\Psi}(\omega_1)+q_2\bar{\Psi}(\omega_2)-q\bar{\Psi}(\omega)
=\frac{\lambda}{8}\tan\frac{\chi}{2}\cot\frac{q\chi}{2}
\cot\frac{q_1\chi}{2}\cot\frac{q_2\chi}{2} \label{eq:ediffq}
\end{equation}
and the right hand side
is negative when $\chi>\chi_q$.  The molecules of size $(q,p)$ dissociate
and the ground state now is a mixture of $\bar{f}_{\omega_1}$ and $\bar{f}_{\omega_2}$ sections with the right proportion.

When $\chi$ continues to increase and reaches $\bar{\chi}_q\equiv \pi/
\max(q_1,q_2)=\min(\chi_{q_1},\chi_{q_2})$, say $\bar{\chi}_q=\chi_{q_1}$,
the molecule of size $(q_1,p_1)$ starts to dissociate into even smaller 
molecules of sizes $(q_2,p_2)$ and $(q_1-q_2,p_1-p_2)$.  Note that 
$(p_1-p_2)/(q_1-q_2)$, $p_1/q_1$, and $p_2/q_2$ are consecutive Farey 
fractions in the $(q-1)$-th level Farey tree.  The process continues until 
all the particles are located at the bottom of the potential well.  This occurs at $\chi>\pi/2$.

It is interesting to observe that for $\chi_q<\chi<\bar{\chi}_q$. even 
though $\bar{f}_{\omega}(x)$ no longer describes  the ground state 
configuration, it remains an increasing function until $\chi$ reaches 
$\bar{\chi}_q$.  Hence molecules of size $(q,p)$ may be regarded as an
unstable resonance state of two smaller molecules of sizes $(q_1,p_1)$ and
$(q_2,p_2)$ when $\chi$ lies in the interval $(\chi_q,\bar{\chi}_q)$.

In summary, for given $\chi$, there exists a positive integer $q_c$ such 
that $\pi/(q_c+1) < \chi \le \pi/q_c$.  The ground state configurations
can only be composed of the $\bar{f}_{\omega}$ sections with $\omega$ in the 
$q_c$-th level Farey tree in the following way.  For an arbitrary irrational 
winding number $\omega$ or rational but not in the $q_c$-th level Farey tree, 
we can find a unique pair of consecutive fractions $\omega_1$ and $\omega_2$ 
in the $q_c$-th level Farey tree such that $\omega_1 < \omega < \omega_2$.  
The ground state configuration with this given $\omega$ can be constructed 
with a fraction ${\bf f_1}$ of particles associated with the 
$\bar{f}_{\omega_1}$ sections and a fraction ${\bf f_2}$ associated with 
$\bar{f}_{\omega_2}$ sections such that
\begin{equation}
{\bf f_1}+{\bf f_2}=1,~~{\bf f_1}\omega_1+{\bf f_2}\omega_2=\omega. \label{eq:frac}
\end{equation}
Moreover, the average energy per particle of this ground state configuration
is given by
\begin{equation}
\bar{\Psi}_e(\omega)=\frac{\omega_2-\omega}{\omega_2-\omega_1}
\bar{\Psi}(\omega_1)+ \frac{\omega-\omega_1}{\omega_2-\omega_1}
\bar{\Psi}_(\omega_2).\label{eq:ein}
\end{equation}

Adding or removing one particle from the system will turn $p_1$ 
$\bar{f}_{\omega_2}$ sections into $p_2$ $\bar{f}_{\omega_1}$ sections or 
vice versa.  This can be seen from Eq.~(\ref{eq:farq}) which implies that
$p_1p_2(1/\omega_1-1/\omega_2)=1$ .  Starting from pure 
$\bar{f}_{\omega_1}$ sections, i.e., a $\bar{f}_{\omega_1}$ configuration,
we can approach a $\bar{f}_{\omega_2}$ configuration by adding particles
one by one.  Solitons and anti-solitons in the usual sense of local
``defects'' do not exists.  Effort to create them merely induces transitions
between $\bar{f}_{\omega_1}$ and $\bar{f}_{\omega_2}$ sections.  If we insist
on calling ``defect'' an $\bar{f}_{\omega_1}$ section in a background of
$\bar{f}_{\omega_2}$ sections, this ``defect'' will carry a ``fractional
charge'' $\frac{1}{p_2}$.

It follows from Eq.~(\ref{eq:ediffq}) that 
\begin{equation}
\left(\frac{q_1}{q_1+q_2}\right)\bar{\Psi}(\omega_1)+\left(\frac{q_2}{q_1
+q_2}\right)\bar{\Psi}(\omega_2)>\bar{\Psi}(\omega)~~{\rm for}~~\chi<\chi_q,
\end{equation}
which indicates that $\bar{\Psi}_e(\omega)$ is a convex function of $\omega$.
Indeed, for given $\chi_{q_c + 1} <\chi \le \chi_{q_c}$, this can be shown by 
making use of Eq.~(\ref{eq:ein}) and proving the 
convexity of $\bar{\Psi}(\omega)$ for any three consecutive fractions 
$\omega_1<\omega_0<\omega_2$ in the $q_c$-th level Farey tree.   
It follows from Eq.~(\ref{eq:farq}) that 
$p_1+p_2 = \kappa p_0$ and $q_1+q_2 = \kappa q_0$ for some positive integer
$\kappa$.  Similar to Eq.~(\ref{eq:ediffq}) one can derive that
\begin{equation}
q_1\bar{\Psi}(\omega_1)+q_2\bar{\Psi}(\omega_2)-\kappa q_0\bar{\Psi}(\omega_0)
=\frac{\lambda}{8}\tan \frac{\chi}{2}\cot\frac{q_0\chi}{2}
\cot\frac{q_1\chi}{2}\cot\frac{q_2\chi}{2}.
\end{equation}
The right hand side is positive and this completes our proof.

Up to now, we have been discussing the ground state configuration for given
$\omega$ as $\chi$ is varied.  Introducing the tensile force term 
$-\sigma\omega$ into Eq.~(\ref{eq:ein}), we obtain the enthalpy of the system.
By minimizing this enthalpy with respect to $\omega$, we obtain the phase 
diagram in the $\chi-\sigma$ plane, as shown in Fig.~1.  One can see that 
there exist tricritical points at $\chi_q=\pi/q$, for each $q$, where three
types of molecules co-exist.  For $\chi<\chi_q$, the lowest enthalpy
configuration is locked to $\omega= p/q$ for 
\begin{equation}
\frac{\bar{\Psi}(\omega)-\bar{\Psi}(\omega_1)}{\omega-\omega_1} \leq 
\sigma \leq  
\frac{\bar{\Psi}(\omega_2)-\bar{\Psi}(\omega)}{\omega_2-\omega},
\end{equation}
where $\omega_1<\omega<\omega_2$ are consecutive fractions in the $q_c$-th
level Farey tree with $\chi_{q_c+1}<\chi\leq \chi_{q_c}$.  When $\chi= 
\chi_q$, the step shrinks to a point and we have 
\begin{equation}
\sigma=\frac{\bar{\Psi}(\omega)-\bar{\Psi}(\omega_1)}{\omega-\omega_1} =\frac{\bar{\Psi}(\omega_2)-\bar{\Psi}(\omega)}{\omega_2-\omega},
\end{equation}
which gives the equations for the tricritical point.  In general, for two
consecutive Farey fractions $\omega_1$ and $\omega_2$, $\sigma= \frac{\bar{\Psi}(\omega_2)-\bar{\Psi}(\omega_1)}{\omega_2-\omega_1}$ gives
the equation for the co-existent curve of the two phases corresponding to
molecules of sizes $(q_1,p_1)$ and $(q_2,p_2)$.

In conclusion, we have shown that the solvable FK model considered in this
work illustrates some important physics accompanying a first order phase
transition.  In particular, we demonstrate explicitly how the Farey tree
dictates the structure of these phase transitions.
Even though the model is one-dimensional and the potential is
not smooth, the underlying physics could very well be realized in nature.

\acknowledgments

The authors would like to express their gratitude to Prof. 
B. Hu for introducing them into this field. 
This work is supported in part by  grants from the National Science
Council  of Taiwan-Republic of China under the contract number
NSC-85-2112-M001-004.

\begin{figure}[h]
\includegraphics{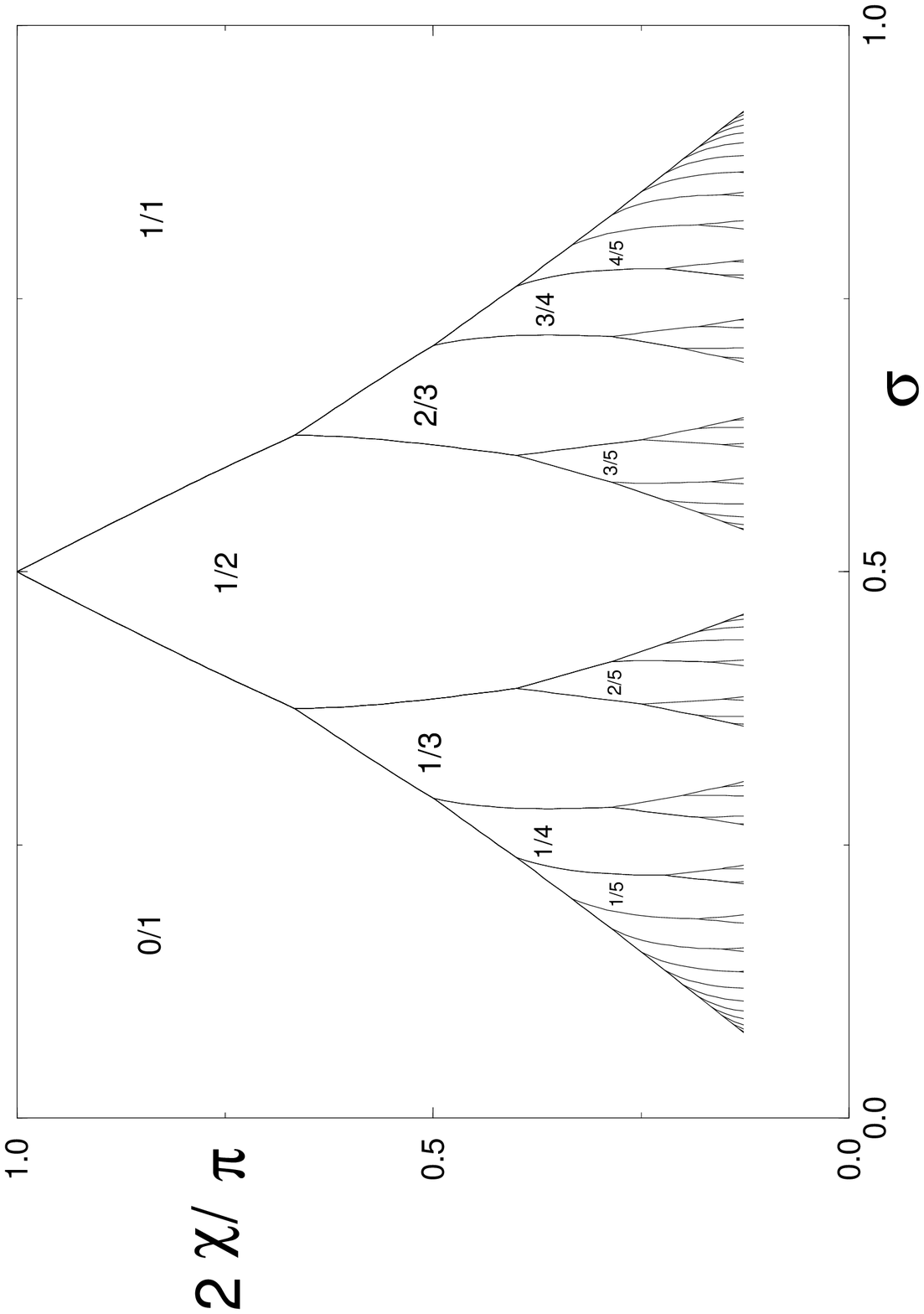}
\vskip 11.0cm
\end{figure}
\centerline{FIG.1. The domains of stability in the $\chi$-$\sigma$ plane.  The 
number in each domain denotes its winding number.}


\begin{references}
\bibitem{irreducible} Throughout this paper, the fraction $p/q$ is in its
irreducible form.
\end{references}
\end{document}